\documentclass[12pt, superscriptaddress, aps,prl,reprint]{revtex4-1}
\usepackage{graphicx}
\usepackage{graphics}
\usepackage{indentfirst}
\usepackage{subfigure}
\usepackage{amsmath, amsthm, amssymb}
\usepackage{color}
\usepackage[normalem]{ulem}
\usepackage{soul}

\newcommand{\phiclus }{\phi^{\rm in}}
\newcommand{\phiout }{\phi^{\rm out}}

\begin{document}

\preprint{This line only printed with preprint option}

\title{Cooperative response and clustering: consequences of membrane-mediated
interactions among mechanosensitive channels}

\author{Lucas D. Fernandes}
\email{ldfernandes@usp.br}
\affiliation{Departamento de Entomologia e Acarologia, Escola Superior de Agricultura "Luiz de Queiroz" - Universidade de S\~{a}o Paulo, ESALQ - USP, 13418-900, Piracicaba/SP, Brazil}
\author{Ksenia Guseva}
\email{ksenia.guseva@uni-oldenburg.de}
\affiliation{Theoretical Physics/Complex Systems, ICBM, University of Oldenburg, 26129 Oldenburg, Germany}
\author{Alessandro P. S. de Moura}
\email{a.moura@abdn.ac.uk}
\affiliation{Institute for Complex Systems and Mathematical Biology, King’s College, University of Aberdeen, AB24 3UE, Aberdeen, United Kingdom}

\begin{abstract}

  Mechanosensitive (MS) channels are ion channels which act as cells' safety
  valves, opening when the osmotic pressure becomes too high and making cells
  avoid damage by releasing ions. They are found on the cellular membrane of a
  large number of organisms.  They interact with each other by means of
  deformations they induce in the membrane. We show that collective dynamics
  arising from the inter-channel interactions lead to first and second-order
  phase transitions in the fraction of open channels in equilibrium relating to
  the formation of channel clusters. We show that this results in a considerable
  delay of the response of cells to osmotic shocks, and to an extreme
  cell-to-cell stochastic variations in their response times, despite the large
  numbers of channels present in each cell.  We discuss how our results are
  relevant for {\it E. coli}.

\end{abstract}

\maketitle

Abrupt changes in the osmolarity of the environment is a hazard most organisms
are subject to at one time or another~\cite{Poolman2004, Wood2001, Roberts2000,
  Wegmann1986, Brown1986, Blomberg1997}. A sudden drop in osmolarity (an
\emph{osmotic shock}) will cause water to rush into a living cell, and requires
an immediate response by the cell to prevent it from getting damaged or
undergoing lysis from the increased tension on the cellular
membrane. \emph{Mechanosensitive channels} (or MS channels) are ion channels
located on the cell membrane, which open when the membrane tension becomes too
high~\cite{Martinac1987, Sukharev1993}, and play a crucial role in the cell's
defence mechanism against osmotic shocks~\cite{Cui1995, Levina1999}. They act as
safety valves, releasing ions and decreasing the osmotic pressure and the
membrane tension. Mechanosensitive channels are found in many organisms, and
have been well characterised in the bacterium \textit{E. coli}~\cite{Kung2005,
  Booth2007, Pivetti2003}.

The cellular membrane in which the mechanosensitive channels are inserted is a
lipid bilayer.  The interior of the bilayer is hydrophobic, making it
energetically favourable for it to thicken or compress to match the hydrophobic
parts of the channel proteins inserted in the membrane~\cite{Phillips2009}. This
results in a deformation profile around each channel, with the thickness of the
bilayer being a function of position.  This deformation mediates a short-range
effective force between two neighbouring channels, similar to the force between
two nearby corks floating on water, which interact through the deformation they
induce on the surface of water. This interaction can be attractive or repulsive,
depending on the shapes of the two molecules.  Furthermore, a theoretical
analysis suggests that the interaction between two neighbouring channels lowers
the tension needed to open them during an osmotic shock~\cite{Ursell2007},
raising the possibility that their function could be influenced by their spatial
distribution on the membrane (as already noticed for other membrane
proteins~\cite{Botelho2006, Skoge2006}).  This is reinforced by the fact that
the channels' attractive forces suggest that they may agglomerate into
clusters. Our goal in this paper is to determine the consequences that the
inter-channel interaction has on the dynamics of this system, focusing in
particular on channel clustering and its consequences for the cell's response to
osmotic shocks.

A preliminary study of mechanosensitive channel clustering was done
in~\cite{Guseva2011, Guseva_book2011}. In that work, diffusion, leading to the
formation of clusters of channels, and opening were considered two separate
processes. This assumption made the model easier to analyse, but it is hard to
justify: in reality, diffusion and gating take place simultaneously.

In this work, we formulate a model of the collective dynamics of
mechanosensitive channels, where diffusion and gating are considered
simultaneous, and no assumption of time separation between clustering and gating
is made.  Using a combination of analytic techniques and numerical simulations,
we analyse the equilibrium and the dynamics of the system, focusing in
particular on the response of the channels to osmotic shocks.  We find that the
interplay between the spatial and the internal degrees of freedom of the
channels leads to unexpected collective phenomena, with possible implications
for their biological function.  We show that the fraction of open channels
undergoes a phase transition as the membrane tension increases; and this
transition changes from second-order to first-order as the density of channels
crosses a critical value.  We explain this change in the nature of the
transition as the result of collective gating induced by a cluster of channels
which appears for high densities.  Studying the time evolution of the system
after applying an osmotic shock, we find that clustering leads to dramatic
changes in the channels' response, slowing down considerably their gating.  In
addition, clustered channels show extreme ensemble variations in their response
times, despite the large numbers of channels present in each cell, what could
translate into large stochastic cell-to-cell differences in response times in a
population of cells.  Finally, we discuss how our results are relevant for the
stress response of \textit{E. coli} and other organisms.


\begin{figure}[ht]
\centering
\includegraphics[width=.45\textwidth]{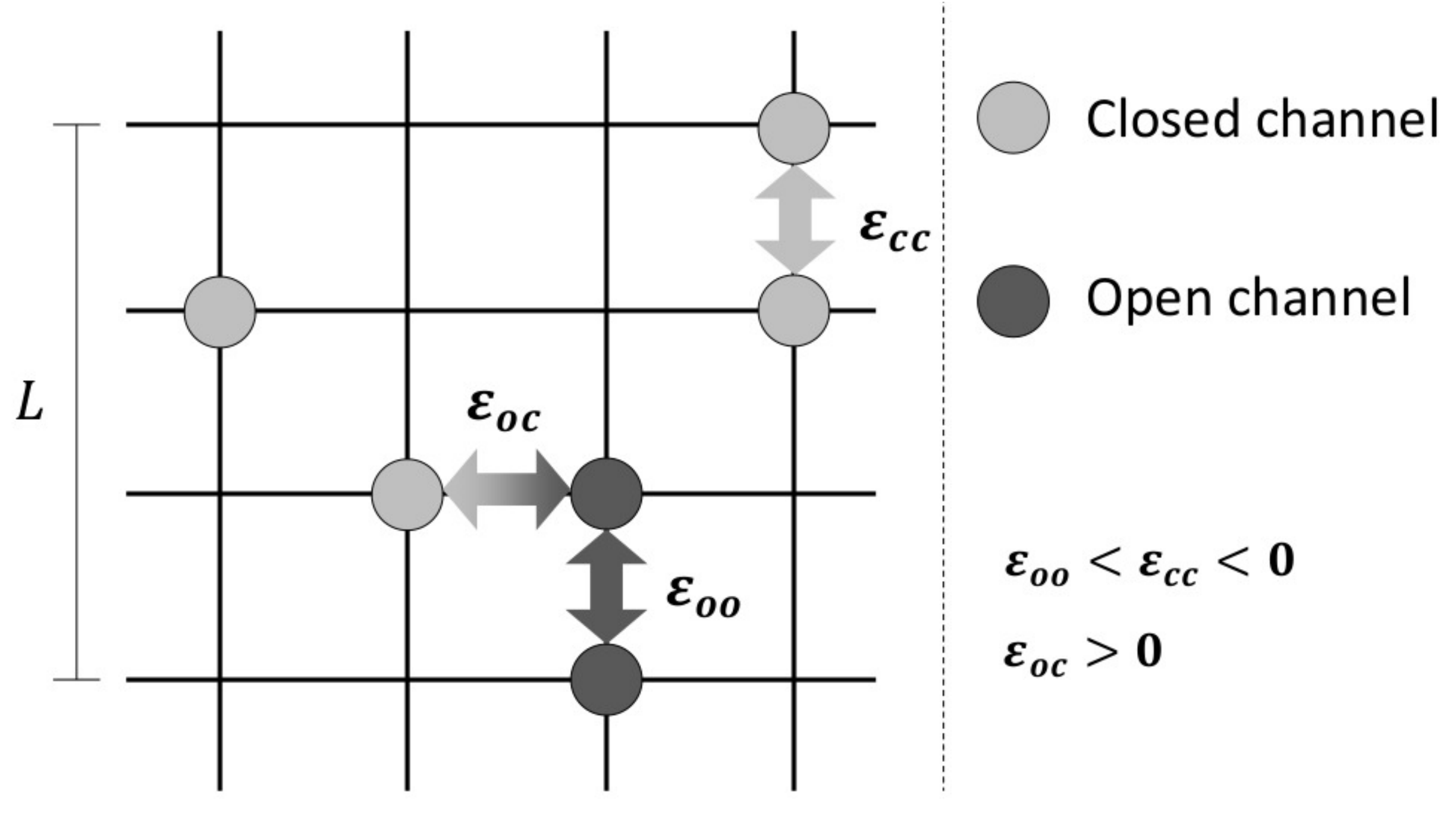}
\caption {Diagram representing the lattice model. Channels interact
  with nearest neighbors with energies $\varepsilon_{oo}$,
  $\varepsilon_{cc}$ and $\varepsilon_{oc}$, depending on the states
  of the interacting channels.}
\label{lattice}
\end{figure}

In our model, we regard the cellular membrane as a two dimensional square
lattice of size $L$, where each of the sites can be either empty or occupied by
a channel.  We focus on the Mechanosensitive Channels of Large Conductance
(MscL), which have been well studied in this context~\cite{Batiza2002,
  Kloda2008, Chang1998}. The total number $N$ of channels is regarded as
constant, so that the density $\rho$ of channels (mean number of channels per
lattice site) is fixed---we are in the canonical ensemble. We consider that MscL
can be in one of two states, closed or open. The interaction energies between
two channels have been obtained by minimising the energy functional defined by
the deformation profile~\cite{Ursell2007}.  For small distances (a few
nanometers) between channels, if $\varepsilon_{oo}$, $\varepsilon_{cc}$ and
$\varepsilon_{oc}$ are the interaction energies between a pair of open channels,
a pair of closed channels, and an open and a closed channel, respectively, we
have $\varepsilon_{oo}<\varepsilon_{cc}<0$ and $\varepsilon_{oc}>0$: two open
channels and two closed channels attract each other, the attraction being
stronger in the first case; and a closed channel repels an open channel
(see~\cite{Guseva2011} for the complete energy profiles). The diagram in
Fig.~\ref{lattice} illustrates our lattice model.

If we consider that all channels are in the same state, without the possibility
of gating, this model is exactly the lattice gas model. The 2D lattice gas model
is exactly solvable, due to its equivalence to the Ising
model~\cite{Newman_book1999}, and presents a phase transition from a homogeneous
to a clustered channel distribution as the density of channels
increases~\cite{Guseva2011}.

The model we describe shares some similarities with the spin-1 Ising model
analysed in a mean-field approximation in the grand canonical ensemble in
\cite{Sivardiere1975}, but our approach has the advantage of yielding more
information about the spatial distribution of channels.


We start by studying the equilibrium properties of the system, as defined by our
lattice model.  We use a mean-field approximation, which will allow us to write
explicit expressions for the energy and entropy of the system, from which we can
find its free energy.  As a simplifying assumption, we assume the existence of
at most one cluster. For the energy values we consider, the existence of a
single cluster in equilibrium is reasonable and supported by test
simulations. Let $f$ be the fraction of channels in the cluster; the other
channels are spread throughout the rest of the membrane. Furthermore, let
$\phiclus$ be the fraction of open channels within the cluster, and $\phiout$
the fraction of open channels outside the cluster.  The three quantities $f$,
$\phiclus$ and $\phiout$ are the thermodynamic variables of our model.  Our next
job is to write the free energy of the system in terms of these variables.
Cluster formation and channel gating are then studied by finding the global
minimum of the free energy.  For example, a cluster is present if $f>0$ in the
state of minimum free energy.

The free energy per channel, $F/N$, for a given temperature $T$, can be written as 

\begin{equation}
 \frac{F}{N} = (e_{int} + e_{mem}) - Ts,
 \label{eq1}
\end{equation}
where the entropy per channel, $s$, can be estimated via combinatorial analysis,
calculating the number of configurations that channels can assume. The energy
per channel is divided into two terms: the interaction among channels,
$e_{int}$, and the interaction of each channel with the membrane, $e_{mem}$. For
each of the configurations devised in the preceding calculation, the interaction
among channels can be estimated considering that channels only interact with
nearest neighbours. The interaction with the membrane depends on the difference
of energies between closed and open states and the work due to the variation on
the channel's area in the gating process.  In the mean-field approximation, we
find (see Supplementary Material for complete derivation of the results):
\begin{align}
 s=&k_B\left\{\ln\left[\frac{(1-\rho)}{\rho(1-f)(1-\phiout )}\right]+ \right. \notag \\
 &\left. f\ln\left[\frac{\rho (1-f)(1-\phiout )}{(1-\rho f)(1-\phiclus )}\right]+ \frac{1}{\rho}\ln\left[\frac{(1-\rho f)}{(1-\rho)}\right]+ \right. \notag \\
 & \left.  \phiout (1-f)\ln\left[\frac{(1-\phiout )}{\phiout }\right] + 
   \phiclus f\ln\left[\frac{(1-\phiclus )}{\phiclus }\right] \right\};
    \label{eq_s}
\end{align}
\begin{align}
 e_{int}=2f(\varepsilon_{cc}+2(\varepsilon_{co}-\varepsilon_{cc})\phiclus + \notag \\
 \quad +(\varepsilon_{cc}-2\varepsilon_{co}+\varepsilon_{oo})(\phiclus )^2);
  \label{eq_eint}
\end{align}
\begin{align}
 e_{mem}=\frac{(\Delta G_0 - \tau\Delta A)}{2}(2f\phiclus +2(1-f)\phiout -1).
  \label{eq_emem}
\end{align}
Here the parameters $\Delta G_0$, $\Delta A$ and $\tau$ are the difference
between the energies of open and closed states, the difference in membrane areas
between the open and closed configurations of a channel, and the membrane
tension, respectively. The term $\Delta G_0$ covers both the energetic cost of
membrane deformation and the cost of changing the internal structure of the
channel. We used $\Delta G_0=50\ k_BT$ and $\Delta A = 20\ \mbox{nm}^2$,
following~\cite{Chiang2004}. For these parameters, a single non-interacting
channel has a 50\% opening probability at the tension
$\tau=2.5\ k_bT/nm^2$~\cite{Ursell2007}.

\begin{figure}[ht]
\centering
\includegraphics[width=.45\textwidth]{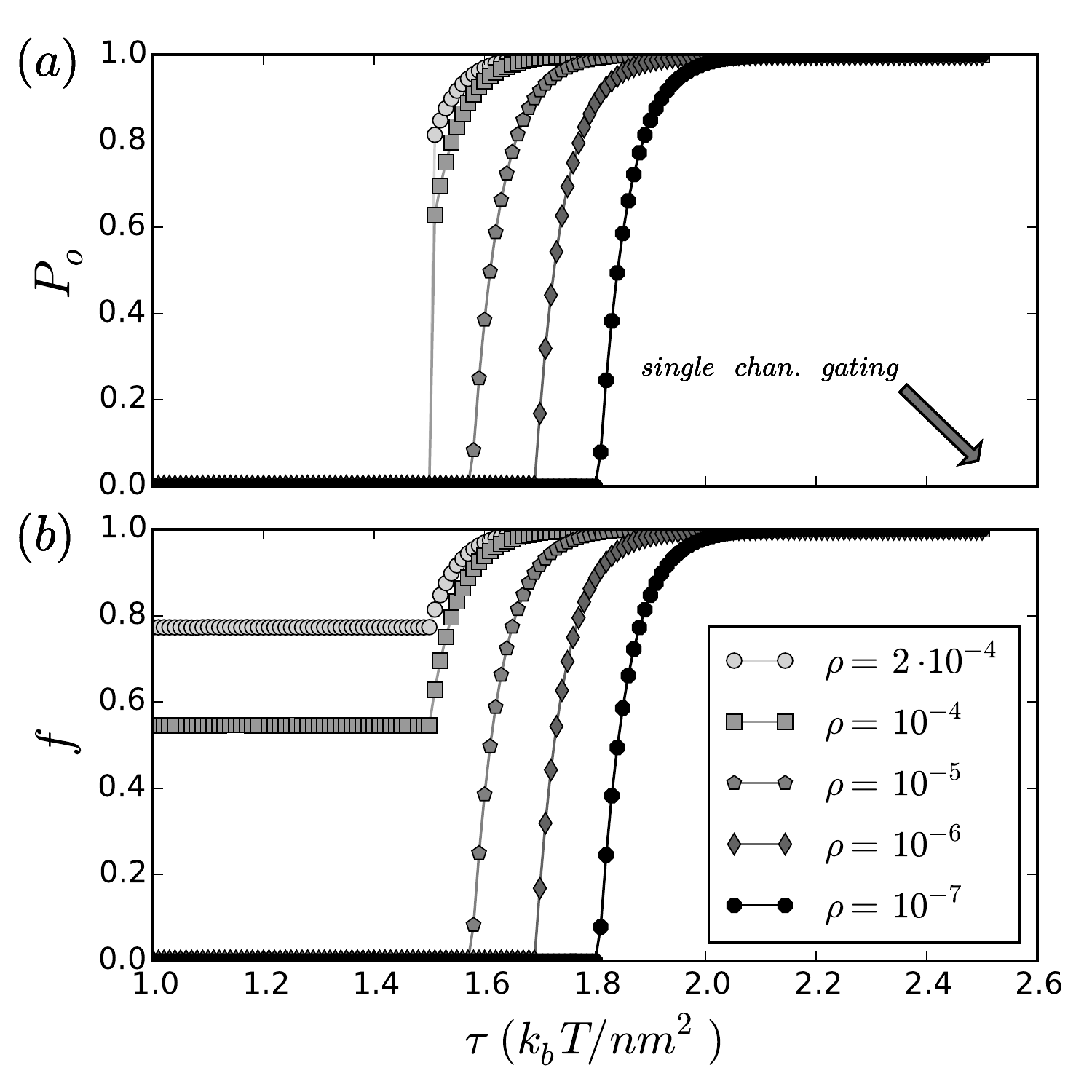}
\caption {Fraction of open channels in the system, $P_o$ (a), and fraction 
of channels that belong to the cluster, $f$ (b), as functions of the membrane
  tension $\tau$, for different values of $\rho$. Here
  $\varepsilon_{oo}=-15.0\ k_BT$, $\varepsilon_{cc}=-5.0\ k_BT$ and
  $\varepsilon_{oc}=10.0\ k_BT$.}
\label{fig1}
\end{figure}

The equilibrium distribution for this system is then given by the values
$f_{eq}$, $\phiout_{eq}$ and $\phiclus_{eq}$ which minimize the free energy, for
given values of $\rho$ and $\tau$ (in the following, the subscript {\it eq} will
be omitted).  The fraction of open channels on the whole lattice is given by
$P_o=f\phiclus +(1-f)\phiout$. Figure~\ref{fig1} shows how $f$ and $P_o$ vary as
functions of the membrane tension, $\tau$, for different values of the density
$\rho$.

We see in Fig.~\ref{fig1}(a) that the fraction of open channels $P_o$ undergoes
a transition from nearly zero (all channels closed) to non-zero values as the
membrane tension $\tau$ increases.  The nature of the transition depends on the
channel density $\rho$.  For small values of $\rho$, the transition is
continuous: $P_o$ increases smoothly from zero beyond a critical value of
$\tau$.  For $\rho$ greater than a critical value $\rho_c$, however, the
transition is discontinuous, with $P_o$ jumping abruptly to a positive value at
the critical tension.

The key to explaining this phenomenon is in the corresponding behaviour of the
cluster fraction $f$, depicted in Fig.~\ref{fig1}(b). At low tensions, the
channels are closed.  Since the force between two closed channels is attractive,
they can form a cluster if their density is high enough; this is the case for
the two upper curves in Fig.~\ref{fig1}(b).  Comparing with Fig.~\ref{fig1}(a),
we see that these correspond to the densities for which the transition in $P_o$
is abrupt: if a cluster already exists at low tension, $P_o$ has a discontinuous
transition.  The reason for this comes from the fact that the interaction energy
between two open channels is much greater than any other combination of
channels, and this becomes more and more so as the tension increases, since high
tensions favour the opening of the channels. In equilibrium, if one of the
channels in the cluster is open, all the others are open as well, because any
mixture of open and closed channels incurs a heavy cost in free energy.  So at a
critical tension, the whole cluster opens, and since the cluster contains a
finite fraction of the channels in the cell, this results in the abrupt jump in
$P_o$ seen in Fig.~\ref{fig1}(a).

For lower channel densities, on the other hand, there is no cluster at lower
tensions.  As the tension is increased, it eventually becomes favourable for
channels to open, and as they do, they will tend to bunch together in a cluster,
because of the high open-open interaction energy.  But because in this case
there was no cluster to start with, the number of open channels will increase
gradually as the tension rises, and so will the cluster size.  This predicts
that the cluster size $f$ and the fraction of open channels $P_o$ will undergo a
continuous transition, and increase in tandem.  This is exactly what we see in
Fig.~\ref{fig1}(a).  In both the low-density and high-density regimes, the
clustering reduces considerably the threshold for channel opening---see
Fig.~\ref{fig1}(a)---, which might have implications for the response of the
cell to osmotic shock, as we shall see in the following.  These collective
phenomena are a direct consequence of the inextricable link between the spatial
distribution of channels and their internal gating dynamics.


In order to understand the response of the channels to an osmotic shock, we have
to go beyond the equilibrium theory and look at their time-dependent activation
dynamics.  To study the coupled gating and diffusion dynamics, we use a Monte
Carlo simulation scheme with two possible actions in each step: (i) with
probability $p_G$, a randomly chosen channel attempts changing its state
(closed/open); or, (ii) with probability $1-p_G$, it attempts to move to one of
its four neighbouring sites, if it is vacant. The attempts succeed with a
probability of acceptance, $A$, according to the criterion:
$A = e^{-\beta\Delta E}$ if $\Delta E > 0$, or $A = 1$ if $\Delta E \leq 0$,
where $\Delta E$ is the change in energy between final and initial
configurations of the system following the attempt. Thus, the algorithm is a
variation of the Kawasaki dynamics, for which the position updates are local,
making it suitable for non-equilibrium simulations of the lattice gas
\cite{Newman_book1999}. The probability $p_G$ is determined by the ratio of the
rates of diffusion and gating: $p_G=\lambda_G/(\lambda_G+\lambda_D)$, where
$\lambda_G=1/\Delta t_G$ and $\lambda_D=1/\Delta t_D$ are the rates of gating
and diffusion, given by the experimentally measured characteristic times of
gating and diffusion, $\Delta t_G$ and $\Delta t_D$, respectively. Each Monte
Carlo step is given after $N$ random choices of channels to attempt change of
state or diffusion, where $N$ is the total number of channels. We relate a Monte
Carlo step, $\Delta t_{MC}$, to a real time interval using the weighted average
$\Delta t_{MC}=p_G\Delta t_G + (1-p_G)\Delta t_D$.  In our simulations, we have
used $\Delta t_G = 4\ \mu s$ and
$\Delta t_D =208\ \mu s$~\cite{Booth2014,Kumar2010} (see Supplementary
material), for which we have $\Delta t_{MC}\cong 8\ \mu s$. Since the increase
in channel area during the gating process precludes the determination of a
single value for the lattice constant, we had to choose it in a range of
reasonable biological values. We use $L=400$ and $\rho = 0.002$ for lattice size
and channel density, respectively, in accordance to typical values for {\it
  E. coli} (see Supplementary material). In all our simulations, we start the
system from an equilibrium situation at low membrane tension $\tau$.  We then
increase $\tau$ abruptly, mimicking an osmotic shock, and follow the dynamics of
the channels using the algorithm described above. The value of $\tau$ is kept
fixed throughout the simulation

\begin{figure}[ht]
\centering
\includegraphics[width=.48\textwidth]{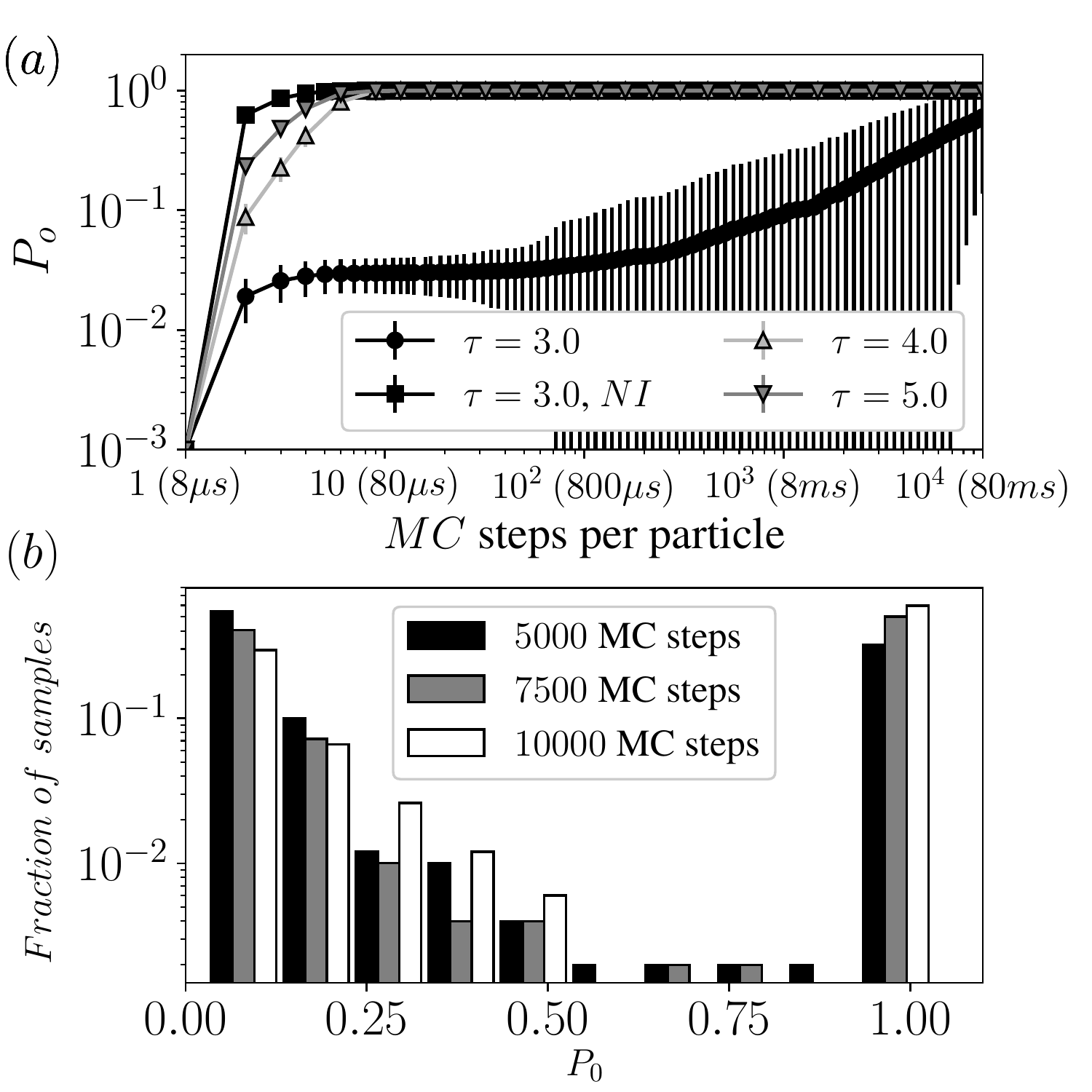}
 \caption {(a) Average fraction of open channels as a function of
   simulated time, for different values of $\tau$. Black squares
   represent the same system without interaction between channels.
   Real time is shown in parentheses on the x-axis. (b) Number of
   samples with a given fraction of open channels, for three different
   transients, for the interacting case with $\tau=3.0\ k_bT/nm^2$.  The
   parameters are: $\lambda_D/\lambda_G=0.02$, $L=400$, $\rho=0.002$, 
   $\varepsilon_{oo}=-15.0\ k_BT$, $\varepsilon_{cc}=-5.0\ k_BT$ and
   $\varepsilon_{oc}=10.0\ k_BT$. For each set of parameters, 500
   samples were considered and error bars correspond to one standard
   deviation.}
\label{fig3}
\end{figure}

It is instructive to compare the time evolution of a hypothetical system of
non-interacting channels to that of the real system of interacting channels.
After approximately $10$ MC steps per channel, all the non-interacting channels
are open and stay in this state until the end of the simulation (black squares,
Fig.~\ref{fig3}a).  The behaviour of the system of interacting channels, in
contrast, is governed by two processes acting on vastly different time scales
(black circles, Fig.~\ref{fig3}a): (i) the fast opening of the isolated channels
outside the cluster; (ii) and the much slower opening of the channels in the
cluster.  The most striking aspect of the dynamics shown in Fig.~\ref{fig3}a is
the dramatic variability of the opening times of the cluster: in one run of the
simulation, the cluster may open in a few microseconds, and in another it may
take $100$ milliseconds to open. This massive variation is a result of the
long-range correlations created by the interactions between channels. The
stochastic nature of the cluster is a direct effect of the nontrivial collective
behavior of the interactive channels.

The variation is further highlighted by the histogram of the fraction of open
channels in a cell some (long) time after the osmotic shock is applied, in $500$
independent runs of the Monte-Carlo simulation (see Fig.~\ref{fig3}b).  We see
that the distribution is bimodal, with roughly similar numbers of cells with
open and closed clusters, even after very long times after the shock.  This
means that in a population of cells subjected to osmotic shock, there will be
massive differences in the response times from one cell to another, even if the
cells are genetically identical and even though they feel exactly the same
stress.  In essence, the collective dynamics that emerged from the channel
interactions amplifies stochastic fluctuations at the molecular scale to the
``macroscopic'', population scale, making them potentially detectable by
population assays.

We note that this large variability disappears once the tension becomes strong
enough. For $\tau = 4.0\ k_bT/nm^2$ and $5.0\ k_bT/nm^2$ (respectively, up and
down-triangles in Fig.~\ref{fig3}a), both clustered and freely-diffusing
channels respond very quickly, with all the channels in the system opening after
only 10 MC steps.

Finally, we would like to emphasize that the membranes are a crowded environment
and that the tight packing of channels withing this environment may introduce
additional effects on gating. The most important effect appears due to the
packing frustration and entropic tension. Packing frustration may lead to a
decrease in the tendency to gate due to space limitation created by
neighbors. Additionally, entropic tension originated by environmental crowding
may have a significant influence on MS channels' conformational change due to
volume exclusion~\cite{Linden2012}.  These effects could be considered, as a
first approximation, accounting to a different choice of parameters in a simple
extension of our model (specifically the energy difference between open and
closed states).

Using fluorescence microscopy and Western blot
analysis~\cite{Bialecka-Fornal2012}, the average number of MscLs in native {\it
  E. coli} cells have been estimated between $300$ and $1000$ channels.  These
results are similar to the one obtained with ribosome profiling~\cite{Li2014}
that measure $360$ to $560$ channels per cell. These numbers take the channel
density close to or above the threshold for cluster formation at low tensions.
Patch-clamp experiments complemented with fluorescent and atomic force
microscopy show evidence for crowding and collective response of channels in
liposomes~\cite{Grage2011}. Other studies have shown non-homogeneous
distributions of overexpressed MS channels in live bacterial cells
~\cite{Wahome2009,Norman2005,Bialecka-Fornal2012}. Although all these studies
suggest cluster formation for native channels, the debate around this question
is still open.  A recent study, through use of PALM (photo-activated
localization microscopy) and SPT (single particle tracking), had shown strong
indications that labeling with fluorescent molecules predisposes MscL channels
to form clusters~\cite{vandenBerg2016}.  In either case, if there is any form of
channel aggregation in bacterial cells, the collective phenomena we describe
here may be directly relevant for the osmotic response of bacteria.  Furthermore
this model can be extended to other types of channels, such as electrically
sensitive ion channels, which are also expected to react cooperatively to
external stimuli~\cite{Cervera20140099}.

Possible evidence for the large variability in channel activation predicted by
our analysis is the recent observation of very late channel gating activity in
\textit{E. coli} cells subject to osmotic shock~\cite{Booth2014}: gating was
seen as long as $100\ ms$ after the shock.  Since isolated MscL channels are
known to gate within a few microseconds after their tension threshold is passed,
it is difficult to explain this observation if the channels do not interact.
This is naturally explained by the variability of channel activation, however:
Fig. \ref{fig3} shows that a cluster could take a time of the order of $100\ ms$
to open. Another recent work~\cite{Buda2016} also highlights the large
cell-to-cell variability of the downshock responses. Furthermore this work shows
a very slow cell volume recovery, which may also indicate channels cooperative
activity.

Calculations based on the ionic flux through single open channels suggest that
as few as $5$ to $10$ channels would be enough to protect a
cell~\cite{Booth2014}.  This contrasts with the recent measurements of MscL
numbers on native cells, which indicate numbers of channels up to two orders of
magnitude greater than this estimate.  With so many channels in a native {\it
  E. coli} cell, simultaneous opening of all channels would lead to a drastic
release of intracellular material, as well as depolarisation of the membrane
potential~\cite{Booth2014}, with potentially fatal consequences for the cell.
Hence this high expression level of channels is still a mystery.  As seen in
Fig.~\ref{fig3}, the presence of the cluster significantly delays the opening of
the whole system of channels, compared with the non-interacting case, for a
shock with smaller membrane tension. Thus, clustering could provide a mean to
self-regulate the simultaneous opening of a large number of channels, in order
to restore the osmotic equilibrium of the cell and function as a channel
reservoir if more of them are needed in case of a severe shock.  This is an
admittedly speculative, but plausible fitness advantage for the large numbers of
channels found in \textit{E. coli}.

\section{Acknowledgments}

It is a pleasure to thank Ian Booth, Heloisa Galbiati and Samantha Miller for
important discussions. This work was supported by the European Union Seventh
Framework Programme [FP7/2007-2013] (NICHE; grant agreement 289384).

\bibliographystyle{apsrev4-1}
\bibliography{mechsen_bib}

\end{document}